\documentclass{article}
\usepackage[utf8]{inputenc}
\usepackage{amsmath,amsthm, amssymb, amsfonts}
\usepackage{fullpage}
\usepackage[pdfencoding=auto]{hyperref}
\usepackage{graphicx}
  \graphicspath{{figures/}}
\usepackage{wasysym}
\usepackage{todonotes}
\usepackage{comment}
\usepackage{mathtools}
\usepackage{wrapfig}
\usepackage{microtype}
\usepackage[algo2e]{algorithm2e}
\usepackage{algorithmic}
\usepackage{footnote}

\usepackage{algorithm,algorithmic}

\usepackage{xcolor}
\usepackage{listings}

\usepackage{color}

\definecolor{codegreen}{rgb}{0,0.6,0}
\definecolor{codegray}{rgb}{0.5,0.5,0.5}
\definecolor{codepurple}{rgb}{0.58,0,0.82}
\definecolor{backcolour}{rgb}{0.95,0.95,0.92}

\lstdefinestyle{mystyle}{
    backgroundcolor=\color{backcolour},
    commentstyle=\color{codegreen},
    keywordstyle=\color{magenta},
    numberstyle=\tiny\color{codegray},
    stringstyle=\color{codepurple},
    basicstyle=\footnotesize,
    breakatwhitespace=false,
    breaklines=true,
    captionpos=b,
    keepspaces=true,
    numbers=left,
    numbersep=5pt,
    showspaces=false,
    showstringspaces=false,
    showtabs=false,
    tabsize=2
}

\usepackage{xparse}

\NewDocumentCommand{\codeword}{v}{%
\texttt{\textcolor{black}{#1}}%
}

\usepackage{xcolor}
\usepackage{tikz}
\usetikzlibrary{3d,decorations.text,shapes.arrows,positioning,fit,backgrounds}
\tikzset{pics/fake box/.style args={
#1 with dimensions #2 and #3 and #4}{code={\draw[gray,ultra thin,fill=#1]  (0,0,0) coordinate(-front-bottom-left) to
++ (0,#3,0) coordinate(-front-top-right) --++
(#2,0,0) coordinate(-front-top-right) --++ (0,-#3,0)
coordinate(-front-bottom-right) -- cycle;
\draw[gray,ultra thin,fill=#1] (0,#3,0)  --++
 (0,0,#4) coordinate(-back-top-left) --++ (#2,0,0)
 coordinate(-back-top-right) --++ (0,0,-#4)  -- cycle;
\draw[gray,ultra thin,fill=#1!80!black] (#2,0,0) --++ (0,0,#4) coordinate(-back-bottom-right)
--++ (0,#3,0) --++ (0,0,-#4) -- cycle;
\path[gray,decorate,decoration={text effects along path,text={CONV}}] (#2/2,{2+(#3-2)/2},0) -- (#2/2,0,0);
}
}}

\tikzset{circle dotted/.style={dash pattern=on 0.05mm off 2mm,line cap=round}}

\allowdisplaybreaks[1]
\usepackage{sistyle} 

\newtheorem{theorem}{Theorem}[section]
\newtheorem*{theorem*}{Theorem}

\newtheorem{definition}[theorem]{Definition}

\newtheorem*{remark*}{Remark}
\newtheorem*{proposition*}{Proposition}

\numberwithin{equation}{section}

\definecolor{darkcandyapplered}{rgb}{0.64, 0.0, 0.0}

\title{{\tt{tfShearlab}}: The TensorFlow Digital Shearlet Transform\\ for Deep Learning}
\author{Héctor Andrade-Loarca\thanks{Institut f\"ur Mathematik, Technische Universit\"at Berlin, 10623 Berlin, Germany, \texttt{andrade@math.tu-berlin.de}} \and Gitta Kutyniok \thanks{Institut f\"ur Mathematik, Technische Universit\"at Berlin, 10623 Berlin, Germany, \texttt{kutyniok@math.tu-berlin.de}} \thanks{Fakult\"at Elektrotechnik und Informatik, 10587 Berlin, Germany} \thanks{Department of Physics and Technology, University of Troms\o, Norway}}

\date{}
\begin{document}
\maketitle

\begin{abstract}

The shearlet transform from applied harmonic analysis is currently the state of the art  when analyzing multidimensional signals with anisotropic singularities. Its optimal sparse approximation properties and its faithful digitalization allow shearlets to be applied 
to different problems from imaging science, such as image denoising, image inpainting, and singularities detection. The shearlet
transform has also be successfully utilized, for instance, as a feature extractor. As such it has been shown to be well suited for image preprocessing in combination with data-driven methods such as deep neural networks. This requires in particular an implementation of the shearlet transform in the current deep learning frameworks, such as TensorFlow.
With this motivation we developed a tensor shearlet transform aiming to provide a faithful TensorFlow implementation. In addition to its usability in predictive models, we also observed an significant improvement in the performance of the transform, with a running time of 
almost 40 times the previous state-of-the-art implementation. In this paper, we will also present several numerical experiments such as image denoising and inpainting, where the TensorFlow version can be shown to outperform previous libraries as well as the learned primal-dual reconstruction method for low dose computed tomography in running time.
\end{abstract}

\medskip
\noindent
\textbf{Keywords:} Software package, deep learning, shearlets, imaging science.
\smallskip

\noindent
\textbf{Mathematics Subject Classification:} 68G05, 68U10, 97N80, 68Q32.

\section{Introduction}

Feature extraction plays an important role in modern image processing and is used as a crucial step in various problem settings of industrial and scientific applications. One example of particular relevance is classification. In fact, currently available approaches typically involve the extraction of features such as edges and blobs, which eases the task of the image classifier \cite{andrade2019wfset,yu2018simultaneous,dalal2005histogram}. The type of features which needs to be extracted certainly depends strongly on the problem one aims to tackle. Blob and line detection, for instance, involve fundamentally different structures \cite{rafael2019symfd,hough1962trans}. Images are often modelled as consisting of features which are delimited by oriented singularities.
Hence these singularities contain an important amount of semantic information in images \cite{torre1986edge,brady1982computational}.
Even more, the human visual system is also known to react most strongly to those curvilinear structures.  For this reason, singularity detection can be considered a key feature extraction task in image processing.

Being able to describe the singularities of an image provides crucial means for efficiently approximating it, usually in terms of what is
referred to as sparse approximation. This task is typically performed by representation systems, which incorporate the specific properties 
of the images in terms of transformations applied to a set of generating functions. In applied harmonic analysis, as a transformation 
leading to different resolutions the action of scaling is customarily applied \cite{meyer1992wavelets}, leading to \textit{multiscale systems}. 
Changing the scale of the generating function enables the system to study features of different sizes, allowing to detect persistent properties along scales such as edges, or more generally, singularities. 
The situation of images as opposed to signals -- i.e., 2D as opposed to 1D data --  requires more attention, since not only the position but also the orientation needs to be encoded by the
representation system. Thus the incorporation of a transformation that changes the orientation becomes necessary, leading to 
\textit{multiscale directional systems}. The system of shearlets introduced in 2006 \cite{gitta2005shearlets} is presumably 
to date the most widely used system of this type. 

The main goal of this paper is to present a digital framework leading to an efficient and faithful implementation of the shearlet transform, which can be utilized directly in the machine learning framework TensorFlow. This allows deployment in various hardware architectures (e.g. GPUs or TPUs), thereby exploiting the inner structure of the operations involved. This is in fact relevant in the current customized machine learning hardware development.

\subsection{Multiscale Directional Systems and Shearlets}

The purpose of applied harmonic analysis is to develop multiscale systems for efficient representations as well as the analysis of 
regularity and detection of features such as singularities. The first celebrated multiscale system were wavelets \cite{mallat1992singularities},
which can be shown to be optimal for singularity extraction in 1D functions. 

As discussed, the 2D case bears additional difficulties in terms of singularities of different {\em orientation}.  
Wavelets are deficient in describing such data due to their isotropic nature. This was formally proven in 2004 by 
Cand\`es and Donoho \cite{candes2004curvelets}. At the same time, they introduced a new multiscale system called
{\em curvelets} by using parabolic scaling and rotation to efficiently represent curvilinear singularities.
The system of curvelets can be regarded as a breakthrough in optimal representation of such singularities.
In fact, within the model situation of cartoon-like functions the ability of curvelets to provide optimally sparse
approximation could be proven (cf. \cite{candes2004curvelets}), also constituting the starting point of what is
by now referred to as {\em geometric multiscale analysis}. However, for practical applications, curvelets can not be
faithfully digitalized in the sense of an implementation consistent with the continuum domain theory due to 
the rotation operator, which is a major drawback (see \cite{candes2006discrete}).

To address the problem of a faithful digitalization, several novel multiscale directional systems were introduced
such as  \textit{contourlets} proposed by Do and Vetterli in 2005 \cite{do2005contour}. One can think of this system as a filterbank 
approach to the curvelet transform. The main advantage of contourlets is their availability to provide a faithful digitalization. However, it still suffers from deficiencies in the continuum setting. In fact, both curvelets and contourlets do not provide
a unified treatment of the continuum and digital situation, and/or a theory for compactly supported systems to guarantee 
high spatial localization. This also results in major difficulties to derive straight forward approximation bounds in 
practical applications.

Finally, as an attempt to overcome all those limitations, the \textit{shearlet system} was introduced by one of the authors jointly 
with Labate and Guo in 2006 \cite{gitta2005shearlets}. This system is generated by few functions, and uses parabolic scaling 
together with a shear operator to change the orientation of its elements. Unlike the rotation operator, the shear operator maintains 
consistency properties with the digital lattice. This now allows to treat the continuum and digital realm uniformly, resulting in a faithful implementation, where most of the theoretical properties are maintained. Also a variant of compactly supported shearlet systems is
available, which was shown to constitute a frame with bounds whithin a numerically stable range \cite{kittipoom2012compact} as well
as provide optimal sparse approximation of cartoon-like functions \cite{kutyniok2011compact}. The shearlet transform as a feature extractor was also theoretically analyzed in terms of resolution of the wavefront set \cite{grohs2011continuous, kutyniok2009resolution}.
The wavefront set of a distribution is a notion from microlocal analysis and contains the position of the singularities of the 
distribution and their respective orientation.

Several implementations of shearlets are available. Maybe the most favorable one for applications due to its speed and accuracy
is the implementation of the  compactly supported shearlets with a flexible shearing operator \cite{kutyniok2016shearlab}. 
The current code is available on {\tt{www.shearLab.org}}.

\subsection{Feature Maps and Convolutional Neural Networks}
\label{SecFeat}

Deep neural networks have played a major role in the last few years in both research and industry, with its significance 
predicted to even increase in the future. A {\em neural network} is a non-linear model made of sequential applications of 
an affine transformation encoded by weights together with a non-linearity. The weights of the network are fitted or 
learned by minimizing a loss function over a training data set. The affine transformation allows to map the input 
to a convenient representation and the non-linearity is crucial to encode complex patterns in the data set. Finally, the term
{\em deep} refers to a large number of layers corresponding to the number of sequential applications.

The first network architecture that showed state-of-the-art results in an image processing task was the so 
called \textit{LeNet} proposed by Yann LeCun in 1989 \cite{lecun1989lenet}. This network was able to classify 
small images of handwritten digits with high accuracy. Since the invention of convolutional neural networks, 
neural networks have started a triumphal march with their primary application area still being imaging science. Nowadays, 
convolutional neural networks present state-of-the-art performance in, for instance, image classification 
\cite{krizhevsky2012imagenet} and reconstruction \cite{adler2018learned}.

A {\em convolutional neural network} contains four types of layers:

\begin{itemize}
    \item \textbf{Convolutional layer:} A convolution operator is applied, which results in the input being averaged locally by a filter defined in a small window moving through the entire input. The extension of the window is known as the receptive field of the convolutional layer.
    \item \textbf{Pooling layer:} After the convolutional layer, one performs a dimensional reduction step, also known as pooling. This operation combines outputs of neuron clusters at one layer into a single neuron in the next layer by, for example, taking the average or the maximum.
    \item \textbf{Activation layer:} Similar to all other network architectures, one then applies a non-linear univariate function componentwise
        to the output of the pooling layer. This is key to allowing the learning of complex non-linear patterns in the data. It is known that this layer provides the network translation invariance.
    \item \textbf{Fully connected layer:} In certain situations such as classification, it is also convenient to use fully-connected layers as the last layers.
\end{itemize}

Convolutional networks were originally inspired by the animal visual cortex, where individual cortical neurons respond to stimuli only in a restricted region of the visual field \cite{fukushima1980cortex, hubel1968receptive}, called the receptive field. Besides its biological resemblance, it also reduces the number of weights to learn in comparison with networks with fully connected layers. For this reason, these networks are much faster to train. Moreover, the structure of its layers results in shift or space invariance as well as stability under certain (Lipschitz) input deformations \cite{wiatowski2015conv}. This characteristics matches the desiderata of several image processing tasks such as image classification, when the object label is independent to its position.

Due to the convolutional structure, the outputs of each layer can be regarded as those features detected by the particular filters and the extension of the receptive field, e.g., a vertical line or a curve. Which features the network deems to be important is defined during learning, depending on a particular task. Thus each layer constitutes a feature map. In this sense a convolutional neural network can be regarded as a sequential and hierarchical application of feature extractors. Those feature maps will presumably represent different types of features varying on size, direction, and amplitude depending on the task at hand. 

Given training data, a convolutional neural network learns those filters (or feature maps) which in a certain sense best matches the function class represented by this data. Thus such an approach is fully data-driven, i.e., it learns end-to-end a suitable representation of the data for a specific the task.

During the last years, hybrid methods \cite{bubba2018learning,adler2018learned} became popular, since they encode the physics of the problem, reduce the complexity of the learning problem, and allow a better interpretation of the entire pipeline. {\em Hybrid methods} incorporate a-priori information from the task to perform, for example, the underlying data distribution and the function class representing the data, to ease the learning task and at the same time ensure solutions which fit with the desired model.

Since shearlets are known to optimally represent images governed by curvilinear singularities as key features, they can be used as a preprocessing step for a convolutional neural networks in imaging science, thereby also reducing the computational complexity of the respective problem by performing some type of heavy lifting. 
This type of approach has already been used for edge and orientation detection \cite{ben2013wavelets, andrade2019wfset}, showing a significant  advantage over other methods that solve the problem using a pure data-driven approach. In a similar manner, the shearlet transform can be applied to other inverse problems involving deep neural networks, such as computed tomography reconstruction \cite{bubba2018learning}. However, a digital version of the shearlet transform for a widely used machine learning framework such as TensorFlow is clearly in demand. 

\subsection{TensorFlow Scalability and Heterogeneity}

Since its release in 2015, Tensorflow has become the most widely used machine learning framework. Indeed, the community of users and contributors is growing everyday. In coarse terms, TensorFlow is an interface for expressing machine learning algorithms \cite{dean2015tensorflow}. The main advantage over former frameworks such as torch, scikit-learn, caffe, or theano consists in the fact that a computation expressed using TensorFlow can be executed with almost no modification on a wide variety of heterogeous systems, from mobile devices to large-scale distributed systems with a lot of computational devices, such as GPU cards. This system was created by Google Brain as an open-source version of their local framework DistBelief. Google's need on scalability and heterogeneity gave the main inspiration for the framework's design with their main goal being to explore the power of very-large-scale deep neural networks for predictive tasks.

TensorFlow expresses computations associated with a specific algorithm as a dataflow-model, also known as computational graph, and maps them onto the specific hardware architecture. The heterogeneity of the framework, i.e., the possibility to use it in various hardware architectures results from the linear algebra operations compiler XLA (Accelerated Linear Algebra). XLA compiles the computational graph composed by several tensor operations into the specific architecture, making the training and evaluation of the model easily deployed in GPUs and other accelerating chips such as TPUs. Operations such as convolution and fast Fourier transform (fft) are particularly well optimized in TensorFlow and have performance greater than their equivalent in other programming languages.

These facts will also be beneficial for our intended Tensorflow version of the shearlet transform. In fact, to compute shearlet coefficients of an image, convolutions of the shearlet filters with the image need to be preformed making it amenable to the use of the fft-based convolution. Thus, the shearlet transform is indeed very well suited for TensorFlow. In addition, TensorFlow provides full broadcasting capabilities which permit the parallelization of the distinct shearlet coefficients computation. We will show that in exchange it will gain the computational power of this framework.

\subsection{Contributions}

The main contribution of this work is two-fold: We will first derive a representation of the digital shearlet transform as a set of tensor operations, leading ultimately to a faithful implementation in TensorFlow. Second, we will demonstrate the power of implementing signal convolutional transforms using the underlying linear algebra compiler of TensorFlow, and its application to deep learning based solutions.

The main advantages of {\tt{tfShearlab}} are related to the significant improvement in running time in comparison with the other implementations of shearlets, in particular, {\tt{Shearlab.jl}} (\url{www.shearlab.org/software}), with a boost of around 30x in a GPU. Moreover, the heterogeneity of TensorFlow gives this implementation the potential of being deployed in the state-of-the-art accelerating chips, such as TPU, as well as in large-scale clusters. In addition to the performance advantages, as mentioned before, having an implementation of the shearlet transform in TensorFlow, will allow the users to take advantage of the convenient image representation given by the shearlet coefficients to improve their own predictive models.

\subsection{Outline}

This paper is organized as follows. Section \ref{sec:DST} will introduce the main concepts behind the shearlet transform, with its discrete version leading to the \textit{digital shearlet transform}. We will also discuss the associated implementation in the case of compactly supported shearlets. Section \ref{sec:DSTasTO} is devoted to introducing the basic concepts in TensorFlow, namely tensors and their constructors such as the \codeword{matmul} matrix multiplication operator and the Fourier operators \codeword{fft} and \codeword{ifft}. We will then make use of those concepts to write the digital sheralet transform as a tensor operator which requires, in particular, an optimized implementation of the Fourier shift operators \codeword{fftshift} and \codeword{ifftshift} using TensorFlow native operations. Section \ref{sec:numerics} contains numerical results and benchmarks of {\tt{tfShearlab}}, in particular, the shearlet decomposition and reconstruction, as well as shearlet based denoising and inpainting. Finally, conclusions and future developements are presented in Section \ref{sec:conclusion}.

\section{Digital Shearlet Transform}
\label{sec:DST}

The goal of this section is to introduce the 2D digital shearlet transform, which operates on 2D arrays such as digital images. We would like to point out that the theory and construction can be extended to the 3D setting to handle videos and general 3D structures. However, this is beyond the scope of this paper, and we refer the interested reader to \cite{kutyniok2016shearlab}. As it is customary in signal and image processing, the construction is often first stated in the continuous case with continuous parameters \cite{gitta2005shearlets}. 

In the sequel we will however immediately focus on the transform which results from discretizing the parameter space while maintaining the continuity in the domain. This yields the discrete shearlet transform generated by the discrete shearlet system, which we now first define.

\begin{definition}[\cite{kutyniok2016shearlab}]
\label{sec2def1}
Let $\varphi, \psi, \tilde{\psi} \in L^2(\mathbb{R}^2)$ and $c=(c_1,c_2)\in (\mathbb{R}_{+})^2$. Then the \emph{shearlet system} $SH(\varphi, \psi, \tilde{\psi}; c)$ is defined by
$$
SH(\varphi, \psi, \tilde{\psi}; c) = \Phi(\varphi; c_1)\cup \Psi(\psi; c) \cup \tilde{\Psi}(\tilde{\psi}; c),
$$
where
$$
\begin{aligned}
\Phi(\varphi; c_1) &= \{ \varphi_m = \varphi(\cdot - c_1 m): m\in \mathbb{Z}^2\},\\
\Psi(\psi; c) &= \{\psi_{j,k,m} = 2^{\frac{3}{4}j}\psi(S_k A_{2^j}\cdot - M_c m): j\geq 0, |k|\leq \lceil 2^{j/2}\rceil, m\in \mathbb{Z}^2 \}, \\
\tilde{\Psi}(\tilde{\psi}; c) &= \{\tilde{\psi}_{j,k,m} = 2^{\frac{3}{4}j}\tilde{\psi}(S_k^{\top}\tilde{A}_{2^j}\cdot - \tilde{M}_cm): j\geq 0, |k|\leq \lceil 2^{j/2}\rceil, m\in \mathbb{Z}^2\},
\end{aligned}
$$
with the parabolic scaling matrices $A_{2^j}$ and $\tilde{A}_{2^j}$ defined by
$$
A_{2^j} = \left(\begin{matrix} 2^j & 0 \\ 0 & 2^{j/2} \end{matrix} \right) \quad \text{and} \quad \tilde{A}_{2^j} = \left( \begin{matrix} 2^{j/2} & 0 \\ 0 & 2^j \end{matrix} \right),
$$
the shearing matrix $S_k$ is given by
$$
S_k = \left(\begin{matrix} 1 & k \\ 0 & 1 \end{matrix}\right),  
$$
and the translation matrices $M_c$ and $\tilde{M}_c$ are defined as
$$
M_c \left( \begin{matrix} c_1 & 0 \\ 0 & c_2 \end{matrix} \right) \quad \text{and} \quad \tilde{M}_c = \left(\begin{matrix} c_2 & 0 \\ 0 & c_1 \end{matrix} \right),
$$
respectively.
\end{definition}

The purpose of the parabolic scaling matrices is to change the size of the specific shearlet anisotropically, thereby aiming to model anisotropic features at different scales, where the shearing matrix changes the orientation to capture features at different directions.

Using the notion of discrete shearlet system as stated, we can now define the discrete shearlet transform.

\begin{definition}[\cite{kutyniok2016shearlab}]
Let $\Lambda = \mathbb{N}_0\times \{ -\lceil 2^{j/2}\rceil,\ldots,\lceil 2^{j/2}\rceil\}\times\mathbb{Z}^2$ be the parameter set. Further, let $SH(\varphi, \psi, \tilde{\psi}; c)$ be a discrete shearlet system and retain the notions from Definiton~\ref{sec2def1}. Then the associated \emph{shearlet transform} of $f\in L^2(\mathbb{R}^2)$ is the mapping defined by
$$
f\longrightarrow SH(\varphi,\psi,\tilde{\psi}) f(m',(j,k,m),(\tilde{j},\tilde{k},\tilde{m}))=(\langle f, \varphi_{m'}\rangle, \langle f, \psi_{j,k,m}\rangle, \langle f, \tilde{\psi}_{\tilde{j},\tilde{k},\tilde{m}}\rangle ),
$$
where $((m',(j,k,m),(\tilde{j},\tilde{k},\tilde{m}))\in \mathbb{Z}^2\times\Lambda\times\Lambda$.
\end{definition}

\subsection{Implementation}

Heading towards a digital version, we next need to discretize the spatial domain, i.e., instead of having generators on $L^2(\mathbb{R}^2)$, we focus on the space $\ell^2(\mathbb{Z}^2)$. We remark that the construction of the digital shearlet transform is inspired by the original construction of the 2D digital wavelet transform \cite{mallat2008wavelets}, paying particular attention to the digitalization of the shearing operator \cite{wangq2010discrete}.

In the sequel, we explain the construction provided in \cite{wangq2010discrete}. For this, let $\psi_1,\varphi_1\in L^2(\mathbb{R})$ be a compactly supported wavelet and an associated scaling function, respectively, satisfying the scale relations
\begin{equation}
\label{sec2eq1}
\varphi_1(x_1) = \sum_{n_1\in\mathbb{Z}}h(n_1)\sqrt{2}\varphi_1(2x_1-n_1)
\end{equation}
and
\begin{equation}
\label{sec2eq2}
\psi_1(x_1) = \sum_{n_1\in\mathbb{Z}}g(n_1)\sqrt{2}\varphi_1(2x_1-n_1).
\end{equation}
In this construction, one assumes the scaling function $\varphi_1\in L^2(\mathbb{R})$ is such that the set
$$
\{\sqrt{2}\varphi(2x-n_1): n_1 \in \mathbb{Z}\}
$$
forms an orthonormal basis for $L^2(\mathbb{R})$, meaning that $h$ and $g$ are given by the coefficient equations
$$
\begin{aligned}
h(n_1) &= \langle \psi_1(x_1),\varphi_1(2x_1-n_1) \rangle
\quad \mbox{and} \quad
g(n_1) &= \langle \varphi_1(x_1),\varphi_1(2x_1-n_1) \rangle 
\end{aligned}.
$$

Next, let $f^{\text{1D}}$ be a function on $\mathbb{R}$ and assume the expansion
$$
f^{1D}= \sum_{n_1\in\mathbb{Z}}f_J^{1D}(n_1)2^{J/2}\varphi_1(2^Jx_1-n_1) \quad 
$$
for fixed, sufficiently large $J>0$, known as the coarsest scale. Let us further assume that $\{h_j(n_1)\}_{n_1\in\mathbb{Z}}$ and $\{g_j(n_1)\}_{n_1\in\mathbb{Z}}$ are the Fourier coefficients of the trigonometric polynomials
$$
\hat{h}_j(\xi_1)= \prod_{k=0}^{j-1}\hat{h}(2^k\xi_1) \quad \text{and} \quad \hat{g}_j(xi_1)=\hat{g}\left(\frac{2^j\xi_1}{2}\right)\hat{h}_{j-1}(\xi_1)
$$
with $\hat{h}_0 \equiv 1$. Using the scaling and wavelet function from \eqref{sec2eq1} and \eqref{sec2eq2}, respectively, one can construct the \textit{non-separable} shearlet generator $\psi$, given by
$$
\hat{\psi} = P(\xi_1/2,\xi_2)\hat{\psi}^{\text{sep}}(\xi),
$$
where $\psi^{\text{sep}}=\psi_1\otimes \varphi_1$ is a separable generating function and the trigonometric polynomial $P$ is a 2D fan filter,   see \cite{do2005fanfilter,kutyniok2016shearlab} for a detailed definition of a fan filter.

By using the relation $S_k A_{2^j} = A_{2^j}S_{k/2^{j/2}}$, we now observe that
$$
\psi_{j,k,m}(\cdot) = \psi_{j,0,m}(S_{k/2^{j/2}}\cdot).
$$
Thus, the discrete shearlet system can be digitalized by discretizing faithfully $\psi_{j,0,m}$ as well as the shear operator $S_{k/2^{j/2}}$.
For simplicity, we are now presenting the explicit form of the discrete shearlet generator and digital shear operator. For a detailed derivation, we refer the reader to \cite{kutyniok2016shearlab}.

We start by studying $\psi_{j,0,m}$ and assume that $\varphi_1$ an orthonormal scaling function, i.e.,
$$
\sum_{n\in\mathbb{Z}^2}|\hat{\varphi}_1(\xi+n)|^2  = 1.
$$
Then, by iterative application of \eqref{sec2eq1} and \eqref{sec2eq2}, the function $\psi_{j,0,m}$ can be expressed as
$$
\hat{\psi}_{j,0,m}(\xi) = 2^{-J}e^{-2\pi i m\cdot A_{2^j}^{-1}\xi} P(A^{-1}_{2^j}Q^{-1}\xi)\hat{g}_{J-j}\otimes \hat{h}_{J-j/2}(2^{-J}\xi)\hat{\varphi}_1(2^{-J}\xi),
$$
where $Q=\text{diag}(2,1)$. If we then assume that $f$  follows the expansion formula
\begin{equation}
\label{sec2scaling}
f(x)= \sum_{n\in\mathbb{Z}^2} f_J(n) 2^{J}(\varphi_1\otimes\varphi_1) (2^Jx_1-n_1,2^Jx_2-n_2),
\end{equation}
we obtain the following representation of the shearlet coefficients:
$$
 \langle f,\psi_{j,0,m}\rangle = (f_J\ast (\overline{p_j\ast W_j}))(A_{2^j}^{-1}2^JM_{c_j}m),
$$
where $p_j(n)$ are the Fourier coefficients of $P(2^{J-j-1}\xi_1,2^{J-j/2}\xi_2)$ and $W_j := g_{J-1}\otimes h_{J-j/2}$.

We next turn to the digitalization of the shear operator $S_{k/2^{j/2}}$, restricting ourselves to merely
stating the definition of the digital shear operator $S^d$. To ensure a faithful digitalization, for each scale $j\in \mathbb{N}_0$,
$S_{2^{-j/2}k}^d$ is defined by
\begin{equation}
\label{sec2eq5}
S_{2^{-j/2}k}^d(f_J) := (((\tilde{f}_J)(S_k\cdot))\ast_1 \overline{h}_{j/2})_{\downarrow 2^{j/2}},
\end{equation}
where $\tilde{f}_J$ is given by
\begin{equation}
\label{sec2eq4}
\tilde{f}_J:= ((f_J)_{\uparrow 2^{j/2}}\ast_1 h_{j/2})
\end{equation}
with $\uparrow 2^{j/2}$ and $\downarrow 2^{j/2}$ denoting the factor $2^{j/2}$ upsampling and downsampling operators, respectively, and $\ast_1$ the convolution in the $x_1$-direction. 

This finally leads to the formal definition of the digital shearlet transform. Notice that we restrict ourselves to present the
digitalization of $\Psi(\psi;c)$, since $\tilde{\Psi}(\tilde{\psi}; c)$ can be digitalized in a similar way and $\Phi(\varphi; c_1)$
can be digitalized as for a wavelet system. 

\begin{definition}[Digital Shearlet Transform \cite{kutyniok2016shearlab}]
\label{sec2def2}
Let $f_J \in \ell^2(\mathbb{Z}^2)$ be the scaling coefficients given in \eqref{sec2scaling}. Then the \emph{digital shearlet transform} associated with $\Psi(\psi;c)$ is defined by
$$
\text{DST}_{j,k,m}^{2D} (f_J) = (\overline{\psi^{d}_{j,k}}\ast f_J)(2^J A_{2^j}^{-1}M_{c_j}m) \quad \text{for}\quad j=0,\ldots, J-1,
$$
where
\begin{equation}
\label{sec2eq6}
\psi_{j,k}^d = S_{k/2^{j/2}}^d(p_j\ast W_j)
\end{equation}
with the shear operator defined by \eqref{sec2eq5} and \eqref{sec2eq4}, $p_j$ and $W_j$ modified as explained above, and the sampling matrix $M_{c_j}$  chosen so that $2^J A_{2^j}^{-1}M_{c_j}m\in \mathbb{Z}^2$.
\end{definition}

In a similar fashion one can define the inverse shearlet transform associated with the dual system $\tilde{\Psi}(\tilde{\psi};c)$  by
\begin{equation}
\label{sec2eq8}
\widetilde{DST^{2D}_{j,k,m}} (f_J) = f_J\ast \overline{\tilde{\psi}^d_{j,k}(m)}
\end{equation}
with normalized dual filters $\tilde{\psi}^d$.
This leads to the reconstruction formula given by 
\[
f_J = (f_J\ast \overline{\varphi}^d)\ast \varphi^d +\sum_{j=0}^{J-1}\sum_{|k|\leq 2^{\lceil j/2 \rceil}} (\text{DST}^{2D}_{j,k,m} (f_J))\ast \gamma^d_{j,k} = \sum_{j=0}^{J-1}\sum_{|k|\leq 2^{\lceil j/2\rceil}} (\widetilde{DST^{2D}_{j,k,m}}(f_J))\ast \tilde{\gamma}^d_{j,k}.
\]

Before turning next to the tensorization of the digital shearlet transform, we would like to point out that the digital Fourier transform is extensively used in the construction of the digital shearlets. This is one key reason which allows to boost the performance of the implementation by using the parallelization power that TensorFlow offers.

\section{Digital Shearlet Transform as Tensor Operator}
\label{sec:DSTasTO}

\subsection{The Tensor Type}

As the name may suggest, TensorFlow is a framework to define and run computations involving tensors. In this context, a {\it tensor} is a multidimensional array, generalizing vectors and matrices to potentially higher dimensions. It is the fundamental abstract type of TensorFlow and has as characteristics a data type and a shape (dimensions), where the data type defines the precision of numerical elements in the tensor, e.g., \codeword{Int32} represents integers up to 32 bits, with a maximum value of $2^{32}-1$. 

Internally, TensorFlow represents tensors as $n-$dimensional arrays of base data types, supporting a variety of element types, including signed and unsigned integers ranging in size from 8 bits to 64 bits, IEEE float and double types, a complex number type, and a string type \cite{dean2015tensorflow}. Typically GPU accelerating devices support at most 32 bits precision \codeword{Float32}. Since lower precision also permits faster computations, typically the user will have to define their tensors choosing that base type.

In TensorFlow, a tensor or \codeword{tf.Tensor} object represents a partially defined array that will eventually produce a value when it is evaluated. In this sense, TensorFlow programs operate by first building a graph of \codeword{tf.Tensor} objects, representing abstractly the pipeline of the computations involving the tensors, and finally running parts of this graph by evaluating the tensors at specific values. One should stress that each of the elements of 
\codeword{tf.Tensors} is of the same known data type, although its shape might be only partially known. This provides the freedom of dynamically defining tensors, which can operator on different data sizes. Hence the shape of a tensor will be fully available after the graph is executed.

TensorFlow allows for certain types of tensors, which will play an specific role in the computations. The main types of those are as follows:

\begin{itemize}
    \item \codeword{tf.Variable}.- The values of this tensor type can be modified across the graph.
    \item \codeword{tf.constant}.- As its name suggests, it is a type which has a fixed value, initially defined.
    \item \codeword{tf.placeholder}.- Inserts a placeholder for a tensor that will be always fed.
    \item \codeword{tf.SparseTensor}.- Represents a sparse multidimensional array as three separate dense tensors, namely \codeword{indices}, \codeword{values}, and \codeword{dense_shape}.
\end{itemize}

\subsection{Tensor Operators}
\label{chapter3subsection2}

As for any predictive model, defining the values of the model parameters is as important as the operations performed over this values. Together, tensors and operations will form a computational graph, an abstract representation of the input-output dataflow defined by the model. For example, in the case of a neural network which classifies images, the input images, the output classes, and the weights will be the tensors in the computational graph, while the matrix multiplication of the weights and the application of the involved non-linearities will be the operations that transform the tensors along the graph. In the sequel, we will refer to the weights together with the operations they perform over tensors as {\it tensor operators}. For example, the matrix together with the multiplication will be the operator related to this matrix. One particularly important operator for our specific application is the 2D fast Fourier transform (fft) operator and its inverse given by \codeword{tf.fft2d} and \codeword{tf.ifft2d} (see also Subsection \ref{subsec:fft}). 

In TensorFlow, the second most important abstract type after the \codeword{tf.Tensor}, are the tensor operations, namely \codeword{tf.Operation}. An \codeword{Operation} is a node in a TensorFlow \codeword{Graph} that takes zero or more \codeword{tf.Tensor} and produces zero or more \codeword{tf.Tensor}. In the standard TensorFlow pipeline, after defining a \codeword{Graph} with tensors and operations, one launches the \codeword{Graph} in a TensorFlow session, which broadcasts the related computation to the specific device one is using. In other words, it uses the abstract dataflow defined by the \codeword{Graph} to evaluate the model in the host hardware. The session is then run by using the command \codeword{tf.Session.run()}.

In the following we state an example of the entire pipeline:

\begin{lstlisting}[language=Python]
>> A = tf.constant([1, 2, 3, 4, 5, 6], shape=[2, 3])
>> B = tf.constant([7, 8, 9, 10, 11, 12], shape=[3, 2])
>> C = tf.matmul(A, B)
>> sess = tf.Session()
>> sess.run(C)
array([[ 58,  64],
       [139, 154]], dtype=int32)
\end{lstlisting}
Going through those steps, first a graph is defined which contains the two main tensors given by
$$
A = \left(\begin{matrix} 1 & 2 & 3 \\ 4 & 5 & 6 \end{matrix} \right)\quad \text{,} \quad B = \left( \begin{matrix} 7 & 8 \\ 9 & 10 \\ 11 & 12 \end{matrix}\right)
$$
as well as one tensor obtained by the tensor operation of matrix multiplication of $A$ and $B$, i.e.,
$$
C = A\cdot B = \left(\begin{matrix} 58 & 64 \\ 139 & 154 \end{matrix} \right).
$$

Finally one should also mention that typically a tensor operator expects the tensor to have the shape \codeword{[?,inner_dim,channels]}, i.e., the first dimension is typically of undefined size and stacks different data samples (for example the batch size for training), the inner dimensions will then be the spatial dimension of the represented array (for example, the resolution of an image), and the last dimension corresponds to the channels of the tensor (for example, the RGB channels of a color image).

\subsection{Implementation of the Tensor fftshift and ifftshift in TensorFlow}
\label{subsec:fft}

As mentioned before, our shearlet implementation makes extensive use of the connection between convolutions and Fourier transform to compute the shearlet coefficients. The currently available fast Fourier transform (fft) libraries can in fact exploit different numerical linear algebra techniques to optimally compute the digital Fourier transform of a 2D array, i.e., with minimal running time and memory usage. TensorFlow has its own implementation of fft based on the state-of-the-art FFTW library developed by Frigo and Johnson \cite{frigo1998fft}, callable by the function \codeword{tf.fft}., which we now explain in more detail.

As with other ports of FFTW, in order to save computation time `tf.fft` only computes half of the frequency spectrum, namely the positive frequencies and the null frequency, if the number of samples is odd. The second half of the frequency spectrum, which is the complex conjugate of the first half is just added at the end of this vector. In other words, applying \codeword{tf.fft} results in

\begin{lstlisting}[language=Python]
0 1 2 3 ... (freq bins >0) ... Fs/2 -Fs/2 ... (freq bins < 0) ... -3 -2 -1,
\end{lstlisting}
where \codeword{Fs} is the frequency sample. In order to display a faithful frequency spectrum starting from \codeword{-Fs/2} and ending at \codeword{+Fs/2}, the negative frequency bins need to be shifted. Obviously, applying the inverse Fourier transform requires another shift. 

In other ports of FFTW, such as Matlab or Python, one has access to the functions \codeword{fftshift} and \codeword{ifftshift} that will perform this circular shift on the frequency domain. To illustrate the application of these functions, recall that by convolution theorem that if $f, g\in L^2(\mathbb{R}^2)$, then
\begin{equation}
\label{sec3eq1}
f\ast g = (\hat{f}\cdot \hat{g})^{\vee},
\end{equation}
where $^{\vee}$ represents the inverse Fourier transform. Later, $\hat{g}$ will be a particular shearlet filter in the Fourier domain and $f$ the image one would like to decompose in shearlet expansion, hence $f\ast g$ will be the shearlet coefficient of $f$ associated to the filter $g$.
In pseudocode the digital realization of the forward and inverse Fourier transform of $f$ is
\begin{lstlisting}[language=Python]
Fourier(f)=fftshift(fft(ifftshift(f)))
InverseFourier(f) = fftshift(ifft(ifftshift(f)))
\end{lstlisting}
Using this structure one can implement \eqref{sec3eq1}, and therefore the shearlet decomposition. Similarly, using \eqref{sec2eq8}, the shearlet reconstruction will be implemented using the dual shearlet filters.

This general logic is followed by the Matlab and Julia implementation of {\tt{Shearlab3D}} and can be also used in TensorFlow. However, we face the problem that no Fourier shift functions does exist in TensorFlow. To circumvent this obstacle, we developed a highly efficient implementation of tensor Fourier shift functions, \codeword{tffftshift} and \codeword{itffftshift}, which we now describe: First, we compute the ordering of the indiced using the array processing library \codeword{numpy}, known to be fast for such computational tasks. In the sequel, we will denote the python library \codeword{numpy} by \codeword{np}. Second, we rearrange the target tensor by applying the tensor gathering tool provided by TensorFlow, namely \codeword{tf.gather}. 

To be more detailed, we now provide \codeword{tffftshift} and \codeword{itffftshift} in pseudocode. Let \codeword{xtf} be a \codeword{tf.tensor}, then the Fourier shift can be applied using the following python function:

\begin{lstlisting}[language=Python]
def tfftshift(xtf, axes=None):
    if len(xtf.shape)==3:
        ndim = len(xtf.shape)-1
    else:
        ndim = len(xtf.shape)-2
    if axes is None:
        axes = list(np.array(range(ndim))+1)
    elif isinstance(axes, integer_types):
        axes = (axes,)
    ytf = xtf
    for k in axes:
        n = int(ytf.shape[k])
        p2 = (n+1)//2
        mylist = np.concatenate((np.arange(p2, n), np.arange(p2)))
        ytf = tf.gather(ytf, indices = mylist, axis = k)
    return ytf
\end{lstlisting}
If also \codeword{ytf} is a \codeword{tf.tensor}, the inverse Fourier shift can be performed by the following python function:
\begin{lstlisting}[language=Python]
def itfftshift(ytf, axes=None):
    if len(ytf.shape)==3:
        ndim = len(ytf.shape)-1
    else:
        ndim = len(ytf.shape)-2
    if axes is None:
        axes = list(np.array(range(ndim))+1)
    elif isinstance(axes, integer_types):
        axes = (axes,)
    xtf = ytf
    for k in axes:
        n = int(xtf.shape[k])
        p2 = n-(n+1)//2
        mylist = np.concatenate((np.arange(p2, n), np.arange(p2)))
        xtf = tf.gather(xtf, indices = mylist, axis = k)
    return xtf
\end{lstlisting}

Numerical experiments show that the running time is comparable with the numpy \codeword{fftshift} and \codeword{ifftshift}, which can be considered state-of-the-art. Having these functions in TensorFlow, we are now ready to compute the shearlet coefficients of a tensor.

\subsection{Implementation of {\tt{tfShearlab}}}

The implementation of {\tt{tfShearlab}} is inspired by the julia library {\tt{Shearlab.jl}}, which at the same time exploited ideas from  the original matlab library {\tt{Shearlab3D}} \cite{kutyniok2016shearlab}. It can be divided into three parts:
\begin{itemize}
\item[(1)] Generation of the shearlet filters in Fourier domain until certain scale $J_0$.
\item[(2)] Computation of the shearlet coefficients of an 2D tensor (e.g. an image), also known as the shearlet decomposition.
\item[(3)] Reconstruction of the 2D tensor from the shearlet coefficients, also known as the shearlet reconstruction.
\end{itemize}

Recall that a \codeword{tf.tensor} typically has the shape \codeword{[?,inner_dim,channels]}, allowing broadcasting of operations through different samples stacked in the first dimension, and different channels stacked in the last dimension (cf. Subsection \ref{chapter3subsection2}).

\subsubsection{The Tensor Shearlet Filters}

In our case the shearlet system has different channels representing the distinct shearlet filters for each scale $j \in J := \{0, \ldots, J_0-1\}$ and shearing $k\in K_j := \{ -\lceil 2^{j/2}\rceil, \ldots 0, \ldots, \lceil 2^{j/2}\rceil\}$. Each filter has the same size as the image, namely $(N,M)$. The total number of shearlet filters is then given by 
\begin{equation}
\label{sec3nshearlets}
nShearlets = 2\sum_{j\in J}(|K_j|-1)+1,
\end{equation}
where $|K_j| = 2*\lceil 2^{j/2}\rceil +1$. This implies that the shearlet system in TensorFlow will be a constant tensor with shape
\begin{lstlisting}[language=Python]
>> tf.shape(shearletSystem)
<tf.Tensor 'Const_1:0' shape=(1, N, M, nShearlets) dtype=complex64>
\end{lstlisting}

The shearlet system is computed recursively using a generating wavelet filter $W_j$ and a directional filter $p_j$ as defined in \eqref{sec2eq6}. Although these filters can be chosen almost freely, we provide a default choice, which is inspired by results from \cite{kutyniok2016shearlab}.
More precisely, for the wavelet filter we used a 1D low pass filter $h_{\text{Shearlab}}$ obtained by the Matlab filter design tool using the matlab command
\begin{lstlisting}[language=Matlab]
design(fdesign.lowpass('N,F3dB',8,0.5),'maxflat')
\end{lstlisting}
The 2D filter is then obtained by a tensor product of this filter and its mirror $g_{\text{Shearlab}}$, a high-pass filter. The filter coefficients of this low-pass filter are similar to those of the Cohen-Daubechies-Feauveau (CDF) 9/7 wavelet \cite{cohen1992filter} with four vanishing moments and higher degrees of regularity in the H\"older and Sobolev sense. These properties yield maximal flatness, which is often a superior choice. 

\begin{figure}[htb!]
\centering
\includegraphics[width = 0.25\textwidth]{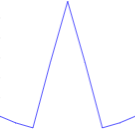}
\hspace*{1cm}
\includegraphics[width = 0.25\textwidth]{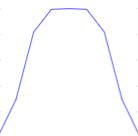}
\hspace*{1cm}
\includegraphics[width = 0.25\textwidth]{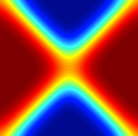}
\put(-372,-12){(a)}
\put(-216,-12){(b)}
\put(-60,-12){(c)}
\caption{(a) The coefficients of the 1D lowpass filter $h_{\text{Shearlab}}$. (b) Magnitude frequency response of $h_{\text{Shearlab}}$. (c) Magnitude response of the 2D fan filter $P_{\text{Shearlab}}$.}
\label{sec3fig1}
\end{figure}

For the directional filter, we make use of the Matlab Nonsubsampled Contourlet Toolbox, obtained with the matlab command
\begin{lstlisting}[language=Matlab]
fftshift(fft2(modulate2(dfilters('dmaxflat4','d')./sqrt(2),'c'))).
\end{lstlisting}
The resulting filter $P_{\text{Shearlab}}$ is a maximally flat 2D fan filter described in \cite{dacunha2006filter}. We refer to Figure \ref{sec3fig1} for an illustration of the coefficients of the low pass filter $h_{\text{Shearlab}}$, its magnitude frequency response, and the magnitude response of the 2D fan directional filter.

The shearlet filters are then computed in the Fourier domain to allow application of the convolution theorem. This approach only requires array multiplications of the corresponding generating filters instead of costly convolutions, followed by scaling in form of upsampling and digital shearing. Those operations are implemented in TensorFlow, enabling to use accelerating hardware such as GPUs to significantly reduce the running time.

In our {\tt{tfShearlab}} API, the function that generates the shearlet system is named \codeword{getshearletsystem2D}. It requires the following parameters:

\begin{itemize}
\item \codeword{row}: The expected horizontal size of the system.
\item \codeword{cols}: The expected vertical size of the system.
\item \codeword{nScales}: The maximum number of scales.
\item \codeword{directionalFilter}: The generating 2D fan directional filter.
\item \codeword{scalingFilter}: The generating 1D low pass filter.
\end{itemize}

The output of this function will be a \codeword{[1, row, cols, nShearlets]} tensor with the shearlet filters, an 1D array with the root mean square of each filter, and an array with the set of corresponding scaling and shearing indexes.

\subsubsection{The Tensor Shearlet Decomposition and Reconstruction}

After generating the shearlet filters, we can compute the digital shearlet transform of any image $I\in \mathbb{R}^{N\times M}$. 
Assuming $J\in \mathbb{N}_0$ scales, the digital shearlet system is given by
$$
DS(N,M,J) = \{ \text{shearletFilters}[i] \in \mathbb{R}^{N\times M}\text{ for } i\in \{1,\ldots,nShearlets(J)\}\},
$$
where $N,M$ are the size in each dimension and $nShearlets(J)$ the number of possible shearlet filters (redundancy) for the maximum sclae $J$, given by \eqref{sec3nshearlets}. The shearlet coefficients will then be computed in Fourier domain as already mentioned in Subsection \ref{subsec:fft}. This leads to the following shearlet decomposition algorithm, where \codeword{Conj} refers to the complex conjugate:
\begin{algorithm}[H]
\hspace*{\algorithmicindent} \textbf{Input:} A digital image $f\in \mathbb{R}^{N\times M}$, a digital shearlet system \codeword{DS(N,M,J)}. \\
 \hspace*{\algorithmicindent} \textbf{Output:} Coefficients \codeword{shearletCoeffs}$\in \mathbb{R}^{N,M,nShearlets}$ \\
 \For{i:=1 to nShearlets do}{
  shearletCoeffs[:,:,i] := fftshift((ifft(ifftshift(shearletFilters[i]$\cdot$Conj((fftshift(fft(ifftshift(f))))))))
 }
 \caption{Shearlet Decomposition \cite{kutyniok2016shearlab}}
 \label{sec3alg1}
\end{algorithm}
Using the TensorFlow functions \codeword{tfftshift}, \codeword{ifftshift}, \codeword{tf.fft2d}, and \codeword{tf.ifft2d}, the tensor shearlet decomposition can then be computed as follows, where \codeword{xtf} is an image tensor of shape \codeword{[1, N, M, 1]} and \codeword{tfshearlets} is the tensor of shearlet filters of shape \codeword{[1, N, M, nShearlets]}:
\begin{lstlisting}[language=Python]
def tfsheardec2D(xtf, tfshearlets):
    xfreqtf = tfftshift(tf.fft2d(itfftshift(xtf)))

    return tfftshift(tf.transpose(tf.ifft2d(tf.transpose(itfftshift(
        tf.multiply(tf.expand_dims(xfreqtf,3),tf.conj(tfshearlets))),
        0,3,1,2])),[0,2,3,1]))
\end{lstlisting}

The shearlet reconstruction, which maps the shearlet coefficients to an image, can be derived similarly by using the dual filters as in \eqref{sec2eq8} and the associated digital dual shearlet system
$$
DualDS(N,M,J) = \{ \text{dualFilters}[i] \in \mathbb{R}^{N\times M}\text{ for } i\in \{1,\ldots,nShearlets(J)\}\},
$$
as follows:
\begin{algorithm}[H]
\hspace*{\algorithmicindent} \textbf{Input:} A set of digital shearlet coefficients \codeword{shearletCoeffs}$\in \mathbb{R}^{N,M,nShearlets}$, a dual digital shearlet system \codeword{DualDS(N,M,J)}. \\
 \hspace*{\algorithmicindent} \textbf{Output:} The reconstructed image $f_{rec}\in \mathbb{R}^{N\times M}$ \\
 // Reconstruction initialization \\
 $f_{rec} := 0 \in \mathbb{R}^{N\times M}$ \\
 \For{i:=1 to nShearlets do}{
  $f_{rec}$ := $f_{rec}$ + fftshift(fft(ifftshift(shearletCoeffs[i])))$\cdot$dualFilters[i]
 }

 $f_{rec}$:= fftshift(ifft(ifftshift($f_{rec}$)))

 \caption{Shearlet reconstruction \cite{kutyniok2016shearlab}}
 \label{sec3alg1*}
\end{algorithm}
We now describe our TensorFlow implementation of Algorithm~\ref{sec3alg1*} in its tensor form, which uses the explicit form of the dual filters from \eqref{sec2eq8}). For this, we let \codeword{coeffstf} be the tensor shearlet coefficients of shape \codeword{[1,N,M,nShearlets]}, \codeword{tfshearlets} the tensor digital shearlet filters of shape \codeword{[1, N, M, nShearlets]}, and \codeword{tfdualFrameWeights} the weights to compute the dual filters from the shearlet filters. The following script is our TensorFlow implementation of Algorithm~\ref{sec3alg2} in its tensor form:
\begin{lstlisting}[language=Python]
def tfshearrec2D(coeffstf, tfshearlets,tfdualFrameWeights ):
    Xfreqtf = tf.reduce_sum(tf.multiply(tfftshift(tf.transpose(
    tf.fft2d(tf.transpose(itfftshift(coeffstf),[0,3,1,2])),[0,2,3,1])),
    tfshearlets),axis=3)

    return tfftshift(tf.ifft2d(itfftshift(tf.multiply(Xfreqtf,
                                1/tfdualFrameWeights))))
\end{lstlisting}
Since this implementation is directly mapped from the Julia API which is based on the faithful formulation of {\tt{Shearlab3D}}, it inherits the theoretical properties of the continuous compactly supported shearlet transform, and even gains significantly in terms of performance improvement and heterogeneity as we will discuss in the next section.

\section{Numerical Results}
\label{sec:numerics}

To test our tensor-based digitalization, we will provide numerical experiments in four different problem settings. More precisely, we will analyze the performance on shearlet decomposition and reconstruction as well as on shearlet-based denoising and inpainting. To also show the potential use of our library with respect to deep learning models for inverse problems, we trained the learned primal-dual algorithm \cite{adler2018learned} for low-dose computed tomography reconstruction on the shearelet coefficients as part of our benchmarks. In order to take advantage of TensorFlow's heterogeneity, we ran our experiments on a graphic card \codeword{Nvidia Quadro GP100} with 16 GB of graphic memory. That this was possible without significant change of the original code gives testament to the expressive power of TensorFlow's computational graphs.

\subsection{Tensor Shearlet Decomposition and Reconstruction}

\begin{figure}[htb!]
\centering
\includegraphics[width = 0.25\textwidth]{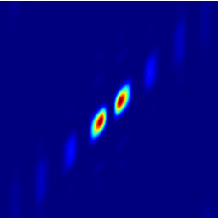}
\hspace*{1cm}
\includegraphics[width = 0.25\textwidth]{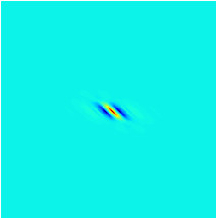}\\[0.3ex]
\includegraphics[width = 0.25\textwidth]{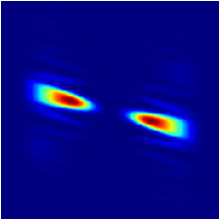}
\hspace*{1cm}
\includegraphics[width = 0.25\textwidth]{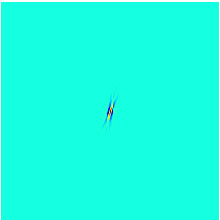}
\put(-216,123){(a)}
\put(-60,123){(b)}
\put(-216,-12){(c)}
\put(-60,-12){(d)}
\caption{Filter corresponding to cone = 1, scale = 1, shear = -2: (a) Frequency domain, (b) Spatial domain; Filter corresponding to cone = 1, scale = 3, shear = 1: (c) Frequency domain, (d) Spatial domain.}
\label{sec4fig1}
\end{figure}

We start with tensor shearlet decomposition and reconstruction. For this, the first step is to compute the corresponding shearlet system.
The {\tt{tfShearlab}} function \codeword{getshearletsystem2D(N, M, nScales, directionalFilter, scalingFilter)} is the constructor of a tensor digital shearlet system with filters of shape \codeword{[N,M]} with maximum number of scales given by \codeword{nScales} and generated with the scaling filter \codeword{scalingFilter} and directional fan filter \codeword{directionalFilter}. Using API \codeword{tfshealab}, the construction can be performed by the following Python commands:
\begin{lstlisting}[language=Python]
# Import the library
import tfshearlab

# Define the parameters
N = 128
M = 128
scalingFilter = 'Shearlab.filt_gen("scaling_shearlet")'
directionalFilter = 'Shearlab.filt_gen("directional_shearlet")'

# Compute the system
tfshearletsystem = tfshearlab.getshearletsystem2D(N,M,nScales,
                   directionalFilter,scalingFilter)
\end{lstlisting}
This will create a computational graph with the operations required to generate the system. To obtain the values, the graph then needs to be evaluated by
\begin{lstlisting}[language=Python]
import tensorflow as tf
tfshearletsystem_values = tf.Session.run(tfshearletsystem)
\end{lstlisting}
This computed system can be applied to images of size \codeword{[N, M] = [128, 128]}. To illustrate the form of the generated filters, we refer to Figures~\ref{sec4fig1}. The construction of this system takes approximately 5.80 seconds, in comparison with the construction of the system in the Julia API, which requires 20.17 seconds. This already shows a 3.5x time improvement. As reasons for this improvement both the use of the GPU and the broadcasting functionality of TensorFlow that optimally vectorize the application of functions can be identified. The main bottleneck is the application of the digital shearing operator which is not optimally implemented in GPU, requiring upsampling and indexing.

\begin{figure}[htb!]
\centering
\includegraphics[width = 0.25\textwidth]{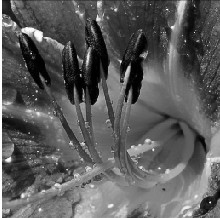}
\caption{Sample image of a flower.}
\label{sec4fig3}
\end{figure}

\begin{figure}[htb!]
\centering
\includegraphics[width = 0.25\textwidth]{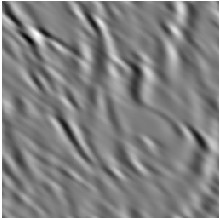}
\hspace*{1cm}
\includegraphics[width = 0.25\textwidth]{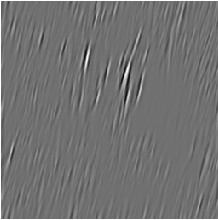}
\put(-216,-12){(a)}
\put(-60,-12){(b)}
\caption{(a) Shearlet coefficients of Figure~\ref{sec4fig3} corresponding to cone = 1, scale = 1 and shear = -2; (b) Shearlet coefficients of Figure~\ref{sec4fig3} corresponding to cone = 1, scale = 3 and shear = 1.}
\label{sec4fig4}
\end{figure}

Having the shearlet filters \codeword{tfshearletsystem} available, we can now compute the shearlet decomposition and reconstruction. As an example image we use the standard flower image depicted in Figure~\ref{sec4fig3}.
After having imported the image in Figure~\ref{sec4fig3} into the python environment using the library \codeword{scipy} by
\begin{lstlisting}[language=Python]
from scipy import ndimage as img
x = img.imread("path-to-image")
\end{lstlisting}
we compute the tensor coefficients of the image with the function \codeword{tfsheardec2D} and evaluate the computational graph
by performing
\begin{lstlisting}[language=Python]
# Convert he image array to tf.Tensor
xtf = tf.constant(x.reshape([1,x.shape[0], x.shape[1]]), dtype= tf.complex64)
# Compute the coefficients
coeffstf = tfshearlab.tfsheardec2D(xtf, tfshearletsystem)
# Evaluate the graph
coeffs = tf.Session.run(coeffstf)
\end{lstlisting}
This operation requires approximately 0.031 seconds. Comparing this with the 1.18 seconds of the standard CPU-based Julia API shows a 30$times$ performance improvement. Figure~\ref{sec4fig4} shows examples of some of the derived shearlet coefficients. 

The shearlet reconstruction can now be performed in a similar manner as the decomposition by using the function \codeword{tfshearrec2D} and running the corresponding computational graph:
\begin{lstlisting}[language=Python]
# Run the reconstruction
xtfrec = tfshearlab.tfshearrec2D(coeffstf, tfshearletsystem)
# Evaluate the graph
xrec = tf.Session.run(xtfrec)
\end{lstlisting}
The running time is 0.033 seconds, whereas the standard CPU-based Julia API requires 1.00 seconds, again showing a 30$\times$ of performance improvement. We remark that -- as already discussed before -- those improvements were expected due to the use of acceleration hardware, such as GPU. In a future we anticipate an even greater improvement with the use of tensor processing units (TPU), which however are so far solely available in the cloud.

\subsection{Shearlet Denoising}

The optimal sparse approximation properties of shearlets make them a perfect candidate for sparse regularization of inverse problems, see, for instance, \cite{genzel2014inpainting}. In this and the next subsection, we will discuss two of such inverse problems, namely denoising and inpainting.

For denoising, we follow the general approach suggested by Labate et al. \cite{labate2012denoising}, which aims to recover the original image from 
its noisy version, where we assume Gaussian white noise. The recovery algorithm is based on hard thresholding the shearlet coefficients of the noisy image. For this, assume that, instead of the original image $f\in \ell^2(\mathbb{Z}^2)$, we only have access to its noisy version, namely
$$
f_{noisy} (i,j) = f(i,j) + e(i,j),
$$
where $e(i,j)\backsim \mathcal{N}(0, \sigma^2)$. We then compute 
$$
f_{denoised} = SH^{-1} \mathcal{T}_{\delta} SH f_{noisy},
$$
where $SH$ denotes the digital shearlet transform and $\mathcal{T}_{\delta}$ the hard thresholding operator given by
$$
(\mathcal{T}_{\delta} x)(n) :=
\begin{cases}
x(n) & \text{if } |x(n)|\geq \delta, \\
0 & \text{else.}
\end{cases}
$$
To increase the performance, we will use a different thresholds $\delta_j$ on each scale $j$, chosen as
$$
\delta_j = K_j \sigma.
$$
In the case of four scales, we typically choose $K = [K_j]_j = [2.5, 2.5, 2.5, 3.8]$.
The quality of the reconstruction will be measured using the peak signal-to-noise ration (PSNR), given by
$$
PSNR = 20 \log_{10}\frac{255\sqrt{N}}{||f-f_{denoised}||_F},
$$
where $N$ is the number of pixels and $||\cdot||_F$ denotes the Frobenius norm. 

In {\tt{tfShearlab}}, denosing is then implemented as follows: 
\begin{lstlisting}[language=Python]
# Compute the shearlet coefficients of the noisy image
coeffstf_noisy = tfshearlab.tfsheardec2D(xtf_noisy, tfshearletsystem)

# Compute the thresholding weights with the RMS of the filters
weights = tf.constant(np.ones(coeffs.shape), dtype = tf.float32)

for j in range(len(tfshearletsystem.RMS)):
    weights[:,:,j] = tfshearletsystem.RMS[j]*tf.constant(np.ones((f.shape[0], f.shape[1])), dtype = tf.float32)

# Hard thresholding
coeffstf = coeffstf_noisy.copy()
zero_indices = tf.abs(coeffs) / (thresholdingFactor * weights * sigma) < 1
coeffstf_noisy[zero_indices] = 0

# Shearlet resconstruction
xtfrec = tfshearlab.tfshearrec2D(coeffstf, tfshearletsystem)

# Run the graph
xrec = tf.Session.run(xtfrec)
\end{lstlisting}
Again studying the running time, for an image of size \codeword{[512, 512]}, the algorithm requires 1.05 seconds with a PSNR of 24.86 dB. For comparable results, the standard CPU-based Julia API takes 1.00 seconds, indicating a 20$\times$ time performance improvement.

\begin{figure}[htb!]
\centering
\includegraphics[width = 0.25\textwidth]{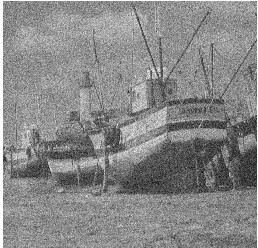}
\hspace*{1cm}
\includegraphics[width = 0.25\textwidth]{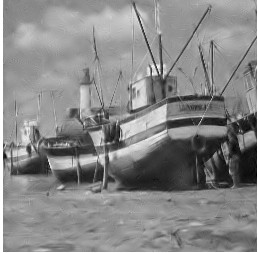}\\[0.3ex]
\includegraphics[width = 0.25\textwidth]{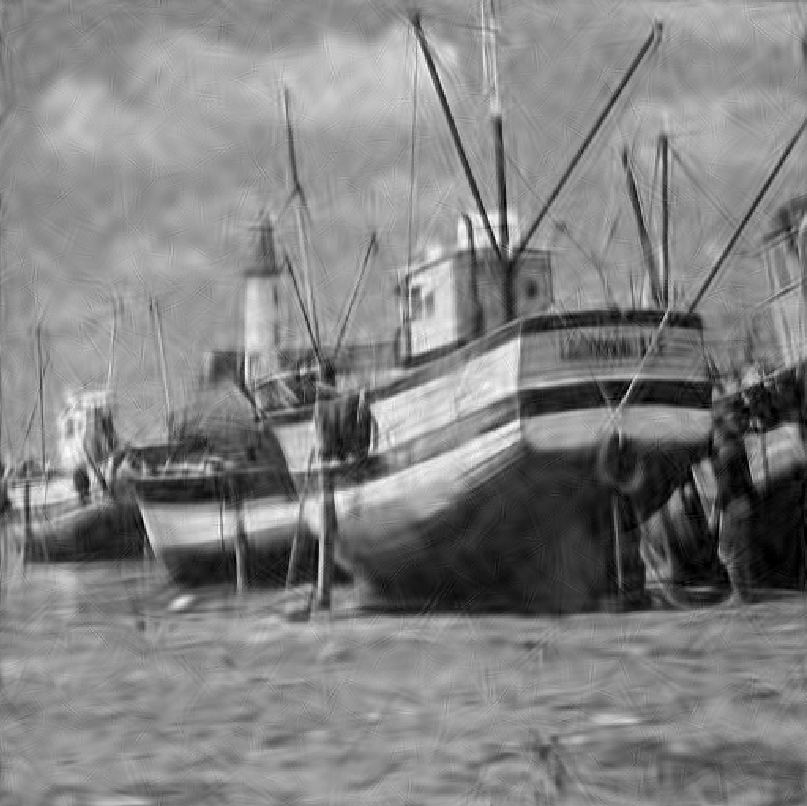}
\hspace*{1cm}
\includegraphics[width = 0.25\textwidth]{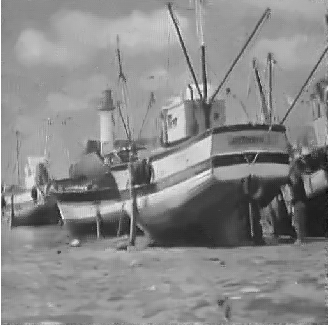}
\put(-216,123){(a)}
\put(-60,123){(b)}
\put(-216,-12){(c)}
\put(-60,-12){(d)}
\caption{Results on denoising: (a) Noisy image, (b) Shearlet-denoised image, (c) Curvelet-denoised image, (d) BM3D-denoised image. }
\label{sec4fig5}
\end{figure}

We would also like to compare with additional standard benchmarks, namely, denosing by curvelets \cite{candes2004curvelets} via thresholding, and the well known Block-matching and 3D filtering algorithm (BM3D) \cite{davob2007bm3d}. We refer to the Table~\ref{table:Denoising-Results} for the obtained PSNR and SSIM on these algorithms. Figure~\ref{sec4fig5} depicts the visual results. In those experiments, we use additive white noise with amplitude of 30. Being superior in running time, we notice that the shearlet denoising approach also performs best concerning accuracy of reconstruction.

\begin{table}[htb!]
\centering
\begin{tabular}{l c c c c}
\textbf{Algorithm} & \textbf{PSNR} (dB) & \textbf{SSIM}\\
\hline
Shearlets thresholding & 26.94 & 0.88\\
\hline

Curvelets thresholding & 26.85 & 0.87 \\
\hline

BM3D & 25.15 & 0.85\\
\hline
\end{tabular}
\caption{Denoising performance of benchmarks for uniform additive noise with amplitude of 30.}
\label{table:Denoising-Results}
\end{table}

\subsection{Shearlet Inpainting}

Inpainting refers to the problem of reconstructing missing parts of an image. In the digital setting, the problem can be formulated as follows: 
A grayscale image $f\in \ell^2(\mathbb{Z}^2)$ is partially occluded by a binary mask $M\in \{0,1\}^{\mathbb{Z}\times\mathbb{Z}}$, i.e.,
$$
f_{masked} (i,j) = f(i,j)M(i,j),
$$
and we aim to reconstruct $f$.

The algorithm for inpainting the missing parts which we will employ is based on an iterative hard thresholding scheme using the sparsifying properties of the shearlet transform \cite{genzel2014inpainting}. In each step a forward shearlet transform is performed on the unoccluded parts of the image combined with the already inpainted features in the missing areas. Then a hard thresholding is applied to the coefficients, followed by the inverse shearlet transform. By gradually decreasing the thresholding constant, this algorithm approximates a sparse set of coefficients whose synthesis approximates the original image on the unoccluded parts. This can be implemented in {\tt{tfShearlab}} as follows:
\begin{algorithm}[H]
\hspace*{\algorithmicindent} \textbf{Input:} $f_{masked}, M, \delta_{init}, \delta_{min}, iterations$ \\
 \hspace*{\algorithmicindent} \textbf{Output:} $f_{inpainted}$\\
 $f_{inpainted} := 0;$\\
 $\delta := \delta_{init};$\\
 $\lambda := (\delta_{min})^{1/(iterations -1)}$\\
 \For{i:=1 to iterations do}{
  $f_{res} := M\cdot (f_{masked}-f_{inpainted});$\\
  $f_{inpainted} := SH^{-1}\mathcal{T}_{\delta}SH(f_{res}+f_{inpainted});$\\
  $\delta:= \lambda\delta$;\\}
 \caption{Shearlet inpainting by iterative hard thresholding}
 \label{sec3alg2}
\end{algorithm}

For our experiments, we used an image of size \codeword{[512,512]} and ran 50 iterations of the iterative hard thresholding for both random and squared mask patterns. The resulting two images showed a PSNR of 24.45 dB and 28.21 dB, respectively. In both cases, the running time was 15 seconds, which is approximately 14$\times$ faster than the classical CPU Julia API.

\begin{figure}[htb!]
\centering
\includegraphics[width = 0.25\textwidth]{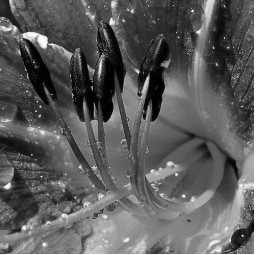}
\hspace*{1cm}
\includegraphics[width = 0.25\textwidth]{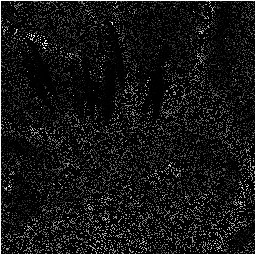} \\[0.3ex]
\includegraphics[width = 0.25\textwidth]{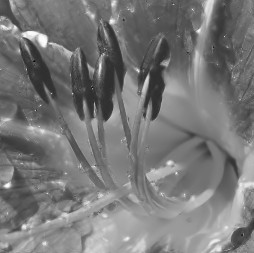}
\hspace*{1cm}
\includegraphics[width = 0.25\textwidth]{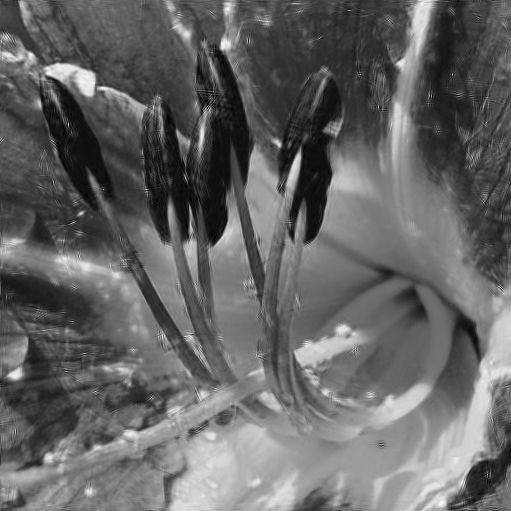}
\put(-216,123){(a)}
\put(-60,123){(b)}
\put(-216,-12){(c)}
\put(-60,-12){(d)}
\caption{Results on inpainting with random mask: (a) The original image, (b) Occluded image, (c) Shearlet-inpainted image, (d) Curvelet-inpainted image.}
\label{sec4fig6}
\end{figure}

\begin{figure}[htb!]
\centering
\includegraphics[width = 0.25\textwidth]{Figures/inpainting/orignal.jpg}
\hspace*{1cm}
\includegraphics[width = 0.25\textwidth]{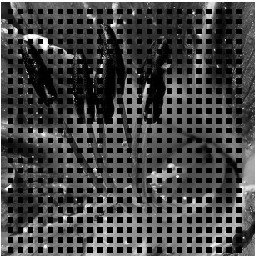} \\[0.3ex]
\includegraphics[width = 0.25\textwidth]{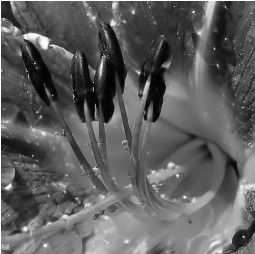}
\hspace*{1cm}
\includegraphics[width = 0.25\textwidth]{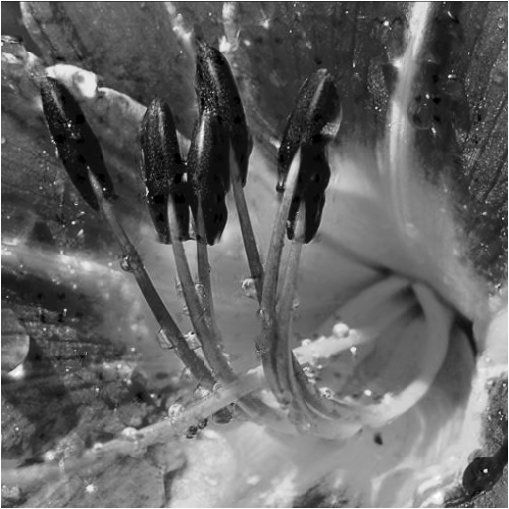}
\put(-216,123){(a)}
\put(-60,123){(b)}
\put(-216,-12){(c)}
\put(-60,-12){(d)}
\caption{Results on inpainting with squared mask: (a) The original image, (b) Occluded image, (c) Shearlet-inpainted image, (d) Curvelet-inpainted image.}
\label{sec4fig7}
\end{figure}

Similar to denoising, we used the same inpainting algorithm but with curvelets as the sparsifying system \cite{candes2004curvelets} as an additional benchmark. We refer to Table~\ref{table:Inpainting-Results} for the obtained PSNR and SSIM of these algorithms. Figures~\ref{sec4fig6} and~\ref{sec4fig7} depict the obtained results for the two used masks. As opposed to the running time, the accuracy of the recovered image is very similar when using shearlets and curvelets. This is not unexpected due to the similarities of both systems.

\begin{table}[htb!]
\centering
\begin{tabular}{l c c c c}
\textbf{Algorithm} & \textbf{Mask} & \textbf{PSNR} (dB) & \textbf{SSIM}\\
\hline
Shearlets inpainting & Random & 28.75 & 0.90\\
\hline
Curvelets inpaitning & Random & 28.60 & 0.90 \\
\hline

Shearlets inpainting & Squared & 32.21 & 0.92\\
\hline

Curvelets inpaitning & Squared & 32.15 & 0.92 \\

BM3D & 25.15 & 0.85\\
\hline
\end{tabular}
\caption{Inpainting performance of benchmarks for random and squared mask.}
\label{table:Inpainting-Results}
\end{table}

\subsection{Learned primal-dual reconstruction on shearlet coefficients}

Inverse problems aim to recover those parameters characterizing the system from observed measurements. In mathematical terms, an
inverse problem can be formulated as follows: Given a forward operator $\mathcal{T}:X\longrightarrow Y$, where $X$ and $Y$ are for now Hilbert spaces, and letting $\delta g\in Y$ be a single sample of a $Y$-valued random variable representing the noise of data $g \in Y$. Then the goal is to recover the true data $f_{\text{true}}\in X$ from measurements $g\in Y$, where
$$
g = \mathcal{T}(f_{\text{true}})+\delta g.
$$

A large class of examples of inverse problems stems from medical imaging applications such as computed tomography (CT). In CT, the forward operator is the Radon transform. The measurement process strikes X-rays through an object coming from a known source. The rays travel on a line $L$ from the source through the object to a detector, being attenuated by the material on the specific line. The precise definition makes this procedure mathematically precise. 

\begin{definition}[Radon transform, \cite{quinto2012introxray}]
Let $f\in L^1(\mathbb{R}^2)$, $\varphi\in [0,\pi)$, and $s\in \mathbb{R}^2$. Moreover, the unit vector in direction $\varphi$ denoted as $\theta$ and its orthogonal vector $\theta^{\perp}$ are given by:
$$
\theta = \theta(\varphi) := (\cos\varphi, \sin\varphi) \quad \theta^{\perp} = \theta^{\perp}(\varphi) := (-\sin\varphi, \cos\varphi).
$$
The \emph{Radon transform} of $f$ is then given by the line integral
\begin{equation}
\label{eq:RayTransform}
\mathcal{R}f(s, \varphi) := \int_{x\in L(\varphi, s)} f(x)dx = \int_{-\infty}^{\infty} f(s\theta(\varphi)+t\theta^{\perp}(\varphi))dt,
\end{equation}
where the line $L(\varphi, s)$ is defined as
$$
L(s, \varphi):= \{ x\in\mathbb{R}^2: x\cdot\theta(\varphi) = s\}.
$$
\end{definition}

In practical cases, often only a limited amount of locations and angles $(s, \varphi)\in \mathbb{R}^2\times [0,\pi)$ can be measured, resulting in highly ill-posed inverse problems. To tackle the problem of reconstruction from CT measurements, Adler and \"Oktem proposed the learned primal-dual reconstruction algorithm \cite{adler2018learned}. This approach exploits neural networks by unrolling a (classical) proximal primal-dual optimization method and using convolutional neural networks instead of the proximal operators. Letting $\mathcal{T}:X\longrightarrow Y$, the classical non-linear learned primal-dual hybrid gradient algorithm (PDHG) also known as the Chambolle-Pock algorithm, can be applied to minimization problems of the form
$$
\min_{f\in X} \left[\mathcal{F}(\mathcal{T}(f))+\mathcal{G}(f)\right],
$$
where $\mathcal{F}:Y\longrightarrow\mathbb{R}$ and $\mathcal{G}:X\longrightarrow \mathbb{R}$ are functionals on the dual/primal spaces.
This algorithm is inspired by the gradient descent algorithm, simultaneously minimizing the primal problem and maximizing the dual problem. In order to generalize the algorithm to also non-smooth functions, in each step the iteration is computed using a proximal operator defined by
$$
\textbf{prox}_{\tau\mathcal{G}}(f) = \text{argmin}_{f'\in X}\left[\mathcal{G}(f')+\frac{1}{2\tau}||f'-f||_X^2\right].
$$
The PDHG algorithm is made precise in the sequel. Notice that $\mathcal{F}^*$ represents the Fenchel conjugate of $\mathcal{F}$, $h\in Y$ is the dual variable, and $\left[\partial\mathcal{T}(f_i)\right]^*:Y\longrightarrow X$ is the adjoint of the derivative of $\mathcal{T}$ at the point $f_i$.
\begin{algorithm}[H]
\hspace*{\algorithmicindent} \textbf{Input:}  $\mathcal{T}$ forward operator, $\sigma, \tau > 0$, s.t. $\sigma\tau||\mathcal{T}||^2 < 1 $, $\gamma \in [0,1]$, $f_0\in X$, $h_0\in Y$ and iterations $I$\\
 \hspace*{\algorithmicindent} \textbf{Output:} $f_{recon}$\\
 \For{i:=1 to $I$ do}{
  $h_{i+1} \leftarrow \textbf{prox}_{\sigma \mathcal{F}^*}(h_i+\sigma \mathcal{T}(\overline{f}_i));$\\
  $f_{i+1} \leftarrow \textbf{prox}_{\tau\mathcal{G}}(f_i-\tau\left[\partial \mathcal{T}(f_i)\right]^*(h_{i+1}))$;\\
  $\overline{f}_{i+1} \leftarrow f_{i+1}+\gamma(f_{i+1}-f_i)$;\\
 }
 \caption{Non-linear primal-dual hybrid gradient algorihtm (PDHG)}
 \label{sec4alg1}
\end{algorithm}

Following the recent trend on hybrid methods combining data-driven and model-based methods, one can take advantage of the power of the structure of the PDHG algorithm, while learning the best update in the primal and dual problem, in some sense of approximating the proximal with neural networks. This idea leads to the learned primal-dual algorithm \cite{adler2018learned}. The approach itself in addition includes minor modifications of the PDHG algorithm such as extending the primal space to allow the algorithm some ``memory'' between iterations, i.e., $f = [f^{(1)}, f^{(2)}, ..., f^{(N_{\text{primal}})}]\in X^{N_{\text{primal}}}$, similarly done in the dual space $Y$, obtaining $Y^{N_{\text{primal}}}$, each $f^{(j)}$ and $Y^{(j)}$ represent an iteration of the step algorithm. To improve the updating step, the network is also allowed to learn how to combine the elements in the evaluation of the proximals.

\begin{algorithm}[H]
\hspace*{\algorithmicindent} \textbf{Input:}  $\mathcal{T}$, $f_0\in X^{N_{\text{primal}}}$ and $h_0\in Y^{N_{\text{dual}}}$ and iterations $I$\\
 \hspace*{\algorithmicindent} \textbf{Output:} $f_{recon}$\\
 \For{i:=1 to $I$ do}{
  $h_i \leftarrow \Gamma_{\theta_i^d}(h_{i-1}, \mathcal{T}(f_{i-1}^{(2)}), g)$;\\
  $f_i \leftarrow \Lambda_{\theta_i^d}\left(f_{i-1}, \left[ \partial \mathcal{T}(f_{i-1}^{(1)})\right]^*(h_{i}^{(1)})\right)$;\\}
 \textbf{return} $f_I^{(1)}$
 \caption{Learned primal-dual algorithm (LPD) \cite{adler2018learned}}
 \label{sec4alg2}
\end{algorithm}

In Algorithm~\ref{sec4alg2}, the operators $\Gamma_{\theta_i^d}$ and $\Lambda_{\theta_i^d}$  approximate the proximal primal and dual operators, known as learned proximal operators, parameterized by $\{\theta_i^d\}\subseteq \Theta$. In the implementation of the learned primal-dual for CT reconstruction \cite{adler2018learned}, the learned proximal operators are residual 3 layered convolutional neural networks of the form
$$
\text{Id}+\mathcal{W}_{w_3,b_3}\circ\mathcal{A}_{c_2}\circ\mathcal{W}_{w_2,b_2}\circ\mathcal{A}_{c_1}\circ\mathcal{W}_{w_1,b_1},
$$
where $W_{w_j, b_j}$ are affine transform with weights $w_j\in X^{n\times m}$ and biases $b_j\in \mathbb{R}^m$ with the $k-th$ component being given by
$$
\left[ \mathcal{W}_{w_j, b_j}([f^{(1)}, ..., f^{(n)})\right]^{(k)} = b_j^{(k)}+\sum_{l=1}^n w_j^{(l,k)}\ast f^{(l)}.
$$
Further, $A_{c_j}$ are non-linearities represented by the Parametric Linear Units (PReLU) functions
$$
A_{c_j}(x) =
\begin{cases}
x & \text{if }x\geq 0, \\
-c_j x & \text{else}.
\end{cases}
$$
In the original implementation, the parameters $N_{\text{primal}}$ and $N_{\text{dual}}$ were set to $5$, the number of iterations $I=10$, the convolutions were of $3\times 3$ pixel size, and the number of channels for each primal learned proximal, $6\rightarrow 32\rightarrow 32\rightarrow 5$, and for the duals $7\rightarrow 32\rightarrow 32\rightarrow 5$. With this setting, the total depth of the network is 60 convolutional layers.

As mentioned in Subsection~\ref{SecFeat} the layers in the convolutional neural networks can be seen as feature maps. In the present situation, we can imagine that these maps aim to to extract key features of the image in the dual and primal space. Having in mind that the shearlet transform is a powerful feature extractor for images, it seems conceivable that performing the learned primal-dual reconstruction in the shearlet coefficients improves its performance. To show that this is indeed the case and providing further evidence of the applicability of {\tt{tfShearlab}} we provide the following three numerical experiments: (1) We trained the learned primal-dual in the image domain, (2) we trained it in the shearlet domain by using the classical ShearLab implementation, and (3) we trained it in the  shearlet domain using the Tensorflow implementation. We used randomly generated ellipses as training set and the Shepp-Loggan Phantom as test set.
In the second and third case, we took each shearlet slice as a channel in the network. This increased the number of parameters of the network by the number of shearlet slices. The implementation of the network was done in TensorFlow, while the ray transform was made using the Python library ODL (Operator Discretization Library) with the ASTRA toolbox as backend. 

We computed the reconstruction for a full angle, low-dose ray transform, where only 30 angles were measured. From the experiments, we observed that the mean squared loss converged faster in terms of iterations when training in the shearlet domain, mainly caused by the convenient representation of the image singularities in the shearlet domain. When the network is trained in the shearlet domain using the classical ShearLab implementation, each iteration takes significantly longer than when trained in the image domain. The main reason of this computational bottleneck is the required streaming of the tensors between the network computations done by Tensorflow in the GPU and the performance of the shearlet transform and its dual done in the CPU. This bottleneck is overcome when using {\tt{tfShearlab}}, due to the reason that then the shearlet coefficients are computed within Tensorflow, removing any need to stream the tensors to the CPU. By using {\tt{tfShearlab}} in combination with the learned primal-dual algorithm, we attained similar performance in terms of training and evaluation time with respect to the non-shearlet version, and improved reconstruction performance.

Table~\ref{table:LPD-Results} shows that the learned primal-dual reconstruction in the shearlet domain achieves better reconstruction performance as compared to training on the image domain -- with performance measured in PSNR and SSIM -- with similar time performance when using {\tt{tfShearlab}}, although significantly slower when using the classical ShearLab implementation. In this table we also depict the performance obtained when training the learned primal-dual approach on the 2D Daubechies 2 (db2) \cite{daubechies1992wavelets} wavelet domain, which presents a significant drop in quality in comparison with the other approaches.
\begin{table}[htb!]
\centering
\begin{tabular}{l c c c c}
\textbf{Model implementation} & \textbf{Training iterations} & \textbf{Time per iteration} & \textbf{PSNR} & \textbf{SSIM}\\
\hline
LPD Image Domain & 50,000 & 0.5 sec & 32.19 & 0.92\\
\hline

LPD Wavelets (2D Daubechies 2) & 20,000 & 0.9 sec & 24.43 & 0.85\\

\hline

LPD Shearlets (classic {\tt{Shearlab}}) & 20,000 & 4.2 sec & 39.19 & 0.98 \\
\hline

LPD Shearlets ({\tt{tfShearlab}}) & 20,000 & 0.8 sec & 39.15 & 0.98\\
\hline
\end{tabular}
\caption{Training and reconstruction performance of the LPD done in image, wavelet domain, and shearlet domain, with classical {\tt{Shearlab}} and {\tt{tfShearlab}}.}
\label{table:LPD-Results}
\end{table}

\begin{figure}[htb!]
\centering
\includegraphics[width = 0.25\textwidth]{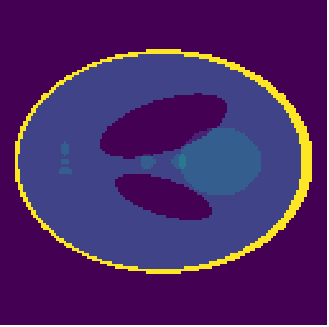}
\hspace*{1cm}
\includegraphics[width = 0.25\textwidth]{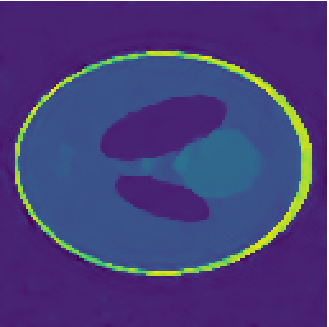}\\[0.3ex]
\includegraphics[width = 0.25\textwidth]{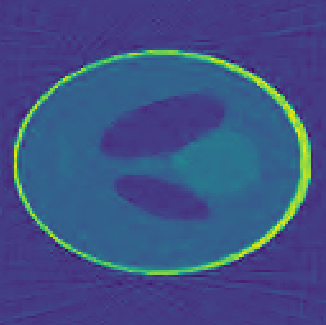}
\hspace*{1cm}
\includegraphics[width = 0.25\textwidth]{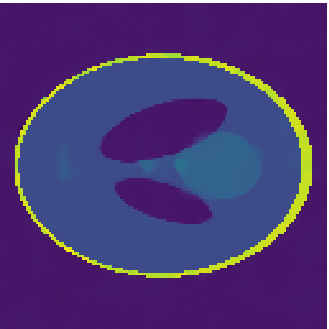}
\put(-216,123){(a)}
\put(-60,123){(b)}
\put(-216,-12){(c)}
\put(-60,-12){(d)}
\caption{Results on CT reconstruction for the Shepp-Logan phantom, when traiend on random ellipses: (a) The Ground Truth, (b) LPD reconstruction on the image domain with 50,000 training steps, (c) LPD reconstruction on the wavelet domain with 20,000 training steps, (d) LPD reconstruction on the shearlet domain with 20,000 training steps.}
\label{fig:lpd-results}
\end{figure}

In Figure~\ref{fig:lpd-results}, one can observe that the learned primal-dual reconstruction on random ellipses achieves a better reconstruction quality evaluated on the Shepp-Logan phantom when it is trained in the shearlet domain, although the quality of both reconstructions is comparable. We also notice that the training in the wavelet domain performs poorly, mainly due to the limitations of wavelets on representing anisotropic features \cite{candes2004curvelets}.

\section{Conclusion}
\label{sec:conclusion}

We observe that the tensor formulation of the shearlet transform is of low cost with the additional advantage of allowing the embedding in different deep learning frameworks, in particular TensorFlow. This is of particular interest for imaging science and computer vision due to the powerful feature extraction and representation of 2D singularities provided by the shearlet transform. In addition to the potential benefits of utilizing the shearlet transform in preprocessing steps of deep learning predictive models, having the shearlet transform at hand in TensorFlow allows to use the heterogeneity and scalability provided by the framework. This significantly improves the performance of the shearlet system generation, decomposition, and reconstruction.

Focusing on the time performance, we observe that the running time of the shearlet decompostion and reconstruction was improved by 30 times as compared to the running time in the previous state-of-the-art implementation, namely, the Julia API of {\tt{Shearlab3D}}. Similar improvements in running time can also be witnessed in the generation of the system, as well as in shearlet denoising and inpainting. The main reason for this superior behavior is the possibility to deploy the algorithms in a graphic card without any change of the underlying code. With the same logic, this method can be deployed without cost on the current state-of-the-art deep learning accelerating hardware, the tensor processing units (TPU). 

In addition to traditional applications of the shearlet transform, we also discussed the application in hybrid algorithmic approaches. We focused on  tomographic reconstruction methods based on deep neural networks, such as the learned primal-dual reconstruction. We showed that {\tt{tfShearlab}} allows to boost the evaluation of the shearlet transform within the network layers. This enables us to take advantage of both the representation power of the shearlet system and the heterogeneity and performance of the tensorflow framework. 

Concluding, the presented tensor version of the shearlet transform provides a highly effective tool that will allow researchers and practitioners to use the shearlet transform in various imaging applications and deep learning models without the necessity of interaction with external frameworks.

\section*{Acknowledgements}

H.A.-L. is supported by the Berlin Mathematical School. 

\bibliographystyle{abbrv}
\bibliography{references}
\end{document}